\documentclass[reprint,showkeys, 
preprintnumbers,amsmath,amssymb]{revtex4}
\usepackage{graphicx}
\usepackage{dcolumn}
\usepackage{bm}
\numberwithin{equation}{section}

\usepackage{hyperref}
\begin{document}
\title{Shifted one-parameter supersymmetric family of quartic asymmetric double-well potentials}
%
\author{Haret C. Rosu}
\email{hcr@ipicyt.edu.mx}
\affiliation{IPICYT, Instituto Potosino de Investigacion Cientifica y Tecnologica,\\
Camino a la presa San Jos\'e 2055, Col. Lomas 4a Secci\'on, 78216 San Luis Potos\'{\i}, S.L.P., Mexico}
\author{Stefan C. Mancas}
\email{mancass@erau.edu}
\affiliation{Department of Mathematics, Embry-Riddle Aeronautical University, Daytona Beach, FL 32114-3900, USA}
\author{Pisin Chen}
\email{pisinchen@phys.ntu.edu.tw}
\affiliation{Leung Center for Cosmology and Particle Astrophysics (LeCosPA) and Department of Physics, National Taiwan University, Taipei 10617, Taiwan\\}

\bigskip

\date{accepted 10 June 2014}
\begin{abstract}
 Extending our previous work ({\em Rosu, Mancas, Chen, Ann.Phys. 343 (2014) 87-102}), we define supersymmetric partner potentials through a particular Riccati solution of the form $F(x)=(x-c)^2-1$, where $c$ is a real shift parameter, and work out the quartic double-well family of one-parameter isospectral potentials obtained by using the corresponding general Riccati solution.
For these parametric double well potentials, we study how the localization properties of the two wells depend on the parameter of the potentials for various values of the shifting parameter.
We also consider the supersymmetric parametric family of the first double-well potential in the Razavy chain of double well potentials corresponding to $F(x)=\frac{1}{2}\sinh 2x-2\frac{(1+\sqrt 2)\sinh 2x}{(1+\sqrt 2) \cosh 2x+1}$, both unshifted and shifted, to test and compare the localization properties.
\end{abstract}
\keywords{Shifted quartic double well, Razavy potential, zero mode, localization, Riccati solution, Darboux deformation.\\
Highlights: Quartic one-parameter DWs with an additional shift parameter are introduced. Anomalous localization feature of their zero modes is confirmed at different shifts. Razavy one-parameter DWs are also introduced and shown not to have this feature.}

\centerline{Ann. Phys. 349 (2014) 33-42}
\centerline{{\small arXiv:1311.6866}}

\maketitle

\section{Introduction}

Anharmonic polynomial potentials, especially the quadratic-quartic double wells (DW), have been used paradigmatically in physics since many years and strengthened our understanding of the natural world at both realistic and fundamental levels. Quadratic-quartic DWs can be encountered in the literature as early as 1918 in the work of Duffing on nonlinear oscillations in such potentials \cite{Duffing}. In our times, dubbed as Mexican-hat potentials, they lie at the core of the origin of mass paradigm for elementary particles according to the Higgs mechanism, and many other important applications can be tracked in physics, such as semiconductor heterostructures, atom transfer in a scanning tunneling microscope, and the Bose-Einstein condensation.

\medskip

In one-dimensional quantum mechanics, one way of introducing DWs is by using directly Riccati equations \cite{Meur} or alternatively the Darboux transformations \cite{Zheng} and simple supersymmetric quantum mechanics \cite{da Silva}. In fact, all these methods are connected mathematically under the common idea of supersymmetry which originally occurred in field theory as a symmetry relating fermions and bosons. A perfect supersymmetry of the world would mean that each elementary particle would have a supersymmetric partner, which is either a fermion or a boson if the original particle is a boson or a fermion, respectively. Except for the spin, all the other quantum numbers of the superpartner particles are the same as well as their mass. Thus, the difference between them might be detected through some peculiar spin interactions which explains why no superpartner particle has been discovered yet. At the level of nonrelativistic quantum mechanics, supersymmetry means that the superpartner potentials are isospectral, i.e., if one knows the spectrum of one of them, the other has identical spectrum except perhaps the ground level and also their eigenfunctions are interrelated. Thus, there is not much hope to distinguish between the two superpartner potentials by some energy-changing processes and for many years people believed that supersymmetric quantum mechanics can be used only to extend the number of exactly solvable quantum problems. Nevertheless, recently there were proposals to look for experimental signatures of the level degeneracy due to supersymmetry in cold atoms in optical lattices \cite{optL1,optL2,optL3}, in arrays of waveguides \cite{Jacobi}, and in tunneling experiments involving a superconducting island coupled to a Josephson junction, the so-called Majorana Cooper-pair box \cite{ulrich}.

However, another interesting class of DWs are the parametric DW potentials obtained in supersymmetric quantum mechanics through general solutions of Riccati equations. Their particular features are that they are related to the Abraham-Moses potentials in inverse scattering and they are strictly isospectral to the potentials from which they are constructed in the sense that even their ground state has the same energy as that of the original potential \cite{M,B1,B2}. In our recent paper dedicated to the parametric supersymmetric potentials \cite{rmc}, we introduced a linear-quadratic-quartic class of one-parameter asymmetric DWs. In the limiting non-parametric case, such potentials occur as effective potentials at the tip of scanning tunneling microscope in the process of atom transfer during a voltage pulse \cite{bg98}. The construction of this type of parametric potentials is based on the general Riccati solution obtained by the well-known Bernoulli ansatz from a particular Riccati solution of the form $F(x)=(x-1)^2-1$. In that case, we noted the interesting feature that the parameter of the potential controls the heights of the localization probability in the two wells, and for certain values of the parameter the height of the localization probability can be higher in the shallower well. In the following the parameter range where this happens will be called {\em anomalous localization region} (ALR), while the remaining range is the regular localization region.

\medskip

The main goal of the present work is to extend the study of these quartic DWs with ALR by employing the same particular quadratic Riccati solutions endowed with an arbitrary real shift parameter $c$, namely $F(x)=(x-c)^2-1$. In other words, we will investigate what happens with the ALR effect when the one-parameter supersymmetric construction is displaced along the axis by $c$ units. In addition, we will use one case of parametric DW from the Razavy chain of supersymmetric potentials to examine the issue of the generality of this interesting localization property.


\section{One-parameter supersymmetric isospectral potentials}

First, we briefly present the mathematical scheme of SUSY QM that we used in \cite{rmc}. The following two Schr\"odinger equations (over the paper, the prime notation is used for the derivatives with respect to $x$)
\begin{equation}\label{e1}
-\Psi''+(V_1-\epsilon)\Psi=0~, \qquad -\tilde{\Psi}''+(V_2-\epsilon)\tilde{\Psi}=0~,
\end{equation}
where $V_1$ and $V_2$ are two different potentials, $\epsilon$ are the spectral eigenvalues, and $\Psi$ and $\tilde{\Psi}$ are the corresponding eigenfunctions, are said to be supersymmetric isospectral partner equations if their potentials satisfy the Riccati equations of the same unknown function
\begin{equation}\label{mh1}
-\Phi'+\Phi^2=V_1-\epsilon~, \qquad \Phi'+\Phi^2=V_2-\epsilon~,
\end{equation}
respectively. In such a case, the spectrum of the second Schr\"odinger equation is identical to that of the first one, i.e., the same $\epsilon$, possibly missing only the ground state.

\medskip

However, if one starts by giving a particular solution, $\Phi_p(x)=F(x)$, of the second Riccati equation in \eqref{mh1}, then  
one can find the one-parameter family of potentials generated by the general Riccati solution sought in the form of the Bernoulli ansatz
\begin{equation}\label{mhG}
\Phi_g(x)=F(x)+ \frac{1}{U(x)}~. 
\end{equation}
Substituting \eqref{mhG} in the second Riccati equation leads to the following first order linear differential equations for $U(x)$
\begin{equation}\label{mh5}
-U'+2F(x)U+1=0~,
\end{equation}
with the solution
\begin{equation}\label{mh6}
U(x)=\frac{\gamma+\int_{0}^x\mu_F(x') dx'}{\mu_F(x)}~, \qquad \mu_F(x)=e^{-2\int_{0}^xF(x')dx'}~,
\end{equation}
where $\mu_F$ is the integrating factor.
Thus
\begin{equation}\label{mh8}
\Phi_g(x)=F(x)+\frac{\mu_F(x)}{\gamma+\int_{0}^x\mu_F(x') dx'}~.
\end{equation}
Using $V_{1\gamma}=V_2-2\Phi_g'$ and $V_1=V_2(x)-2F'(x)$, one immediately gets the one-parameter family of potentials
\begin{equation}\label{mh9}
V_{1\gamma}(x)=V_1(x)-2\left({\rm ln}\left |\gamma-\gamma(x)\right |\right)''~, \qquad \gamma(x)=-\int_{0}^x\mu_F(x')dx'~,
\end{equation}
in which $V_1$ is included for $\gamma=\infty$ and each member of the family has the same supersymmetric partner $V_2$.
The expression $-2\left({\rm ln}\left |\gamma-\gamma(x)\right |\right)''=2\left(\Phi_g^2-\Phi_p^2\right)$ is the parametric Darboux deformation of the potential $V_1$. Depending on how much it deviates from the abscissas line, it gives the local differences between the parametric potential and the original potential $V_1$.
Besides, the unnormalized ground state eigenfunction for each $\gamma$, more generally called zero-mode (ZM), is given by
\begin{equation}\label{mh9b}
\Psi_{0\gamma}(x)= \frac{\sqrt{\mu_F(x)}}{\gamma-\gamma(x)}~. 
\end{equation}
The parameter $\gamma$ defines a range of existence of the family of regular potentials $V_{1\gamma}$ and eigenfunctions $\Psi_{0\gamma}$ which can be obtained graphically. For this, we notice that one can get regular parametric potentials when the lines $\gamma=$ constant do not intersect the graph of $\gamma(x)$.

\medskip

 Parametric potentials of the type (\ref{mh9}) have been long used in physics starting with the work of Mielnik \cite{M} in the particular case of the quantum harmonic oscillator and reviewed by Rosas-Ortiz \cite{rosas98}, and also interpreted as a sequence of two Darboux transformations, see, e.g., \cite{r98}. Most recently, Yang \cite{y93} used examples of these potentials with complex parameter $\gamma$ to illustrate that continuous families of Schr\"odinger solitons cannot bifurcate out from linear guided modes.

 \medskip

 As noted in our previous work \cite{rmc}, for the parametric quartic DW potentials generated by using the Riccati solution $F(x)=(x-1)^2-1$
 the ALR is given by the range of $\gamma\in (\gamma_s,\gamma_{cr}]$, where $\gamma_{cr}$ is the value of $\gamma$ for which the heights of the peaks of the squared ZMs in the two wells become equal. To obtain $\gamma_{cr}$, we used the graphical method for transcendental equations. We noticed that the maxima of $\Psi_{0\gamma}^2$ can be found from the condition $\Phi_g=0$, i.e., from
\begin{equation}\label{14}
-F(x)=\frac{\mu_F(x)}{\gamma^*-\gamma(x)}~, 
\end{equation}
and solving for $\gamma^*$ we get
\begin{equation}\label{15}
\gamma^*=-\frac{\mu_F(x)}{F(x)}+\gamma(x)\equiv \gamma^*(x)~, 
\end{equation}
where we have denoted the r.h.s. of (\ref{15}) by $\gamma^*(x)$. Then,
by intersecting the horizontal lines $\gamma^*=const$ with $\gamma^*(x)$, one can find $\gamma_{cr}$.

 \bigskip



 \section{Shifted quartic parametric DWs}

 Let us take now $F(x)=(x-c)^2-1$. The partner potentials are symmetric about $x=c$ axes and the main quantities are the following ones:
 \begin{align}
& V_{2,1}(x)=\Big((x-c)^2-1\Big)^2\pm 2(x-c)~,\\
 & \mu_F=e^{-\frac{2x}{3}(x^2-3cx+3c^2-3)}~, \\
&V_{1\gamma}(x)=V_1(x)-2\left({\rm ln}\left |\gamma+\int_{0}^x e^{-\frac{2x'}{3}(x'^2-3cx'+3c^2-3)} dx'\right |\right)''~.\\
&\Psi_{0\gamma}(x)= 
\frac{ e^{-\frac{x}{3}(x^2-3cx+3c^2-3)}}{\gamma-\gamma_c(x)}~, \qquad \gamma_c(x)=-\int_{0}^x e^{-\frac{2x'}{3}(x'^2-3cx'+3c^2-3)}~.
\end{align}

\bigskip

For example, for the unshifted case, i.e., $c=0$, we have $F(x)=x^2-1$ and then we get:
\begin{align}
V_{2,1}(x)&=x^4-2x^2\pm 2x+1\equiv (x^2-1)^2\pm 2x~,\\
\mu_F(x)&=e^{-\frac{2}{3}x(x^2-3)}~,\\
V_{1\gamma}(x)&=(x^2-1)^2-2x+\frac{4e^{-\frac{2}{3}(x^2-3)}(x^2-1)}{\gamma-\gamma(x)} 
+\frac{2e^{-\frac{4}{3}(x^2-3)}}{\left(\gamma-\gamma(x)\right)^2}~,\\ 
\Psi_{0\gamma}(x)&=\frac{e^{x-\frac{x^3}{3}}}{\gamma-\gamma(x)}~, \qquad \gamma(x)=-\int_{0}^xe^{-\frac{2}{3}x'(x'^2-3)}dx'~.
\end{align}

\begin{figure}[x!] 
\begin{center}
\resizebox*{0.35\textheight}{!}{
{\includegraphics{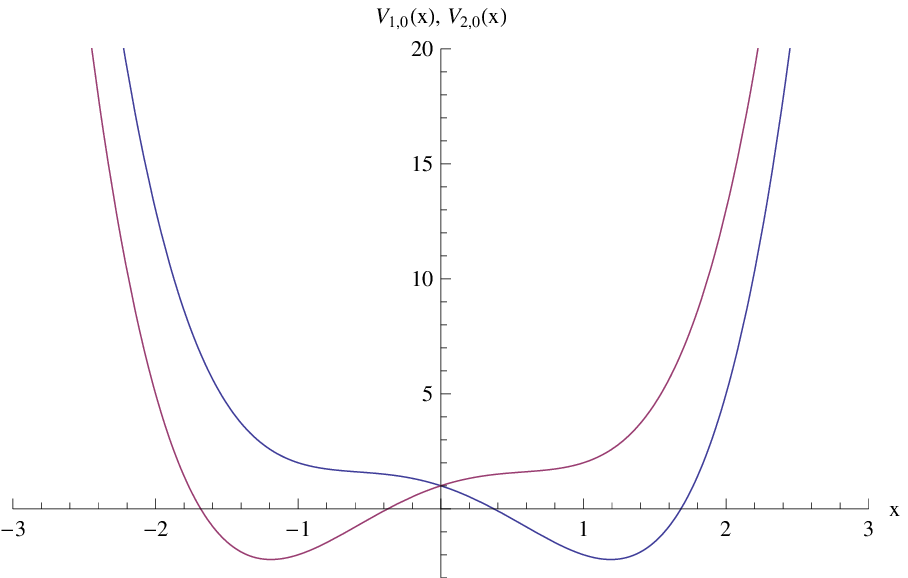}}}
\resizebox*{0.35\textheight}{!}{
{\includegraphics{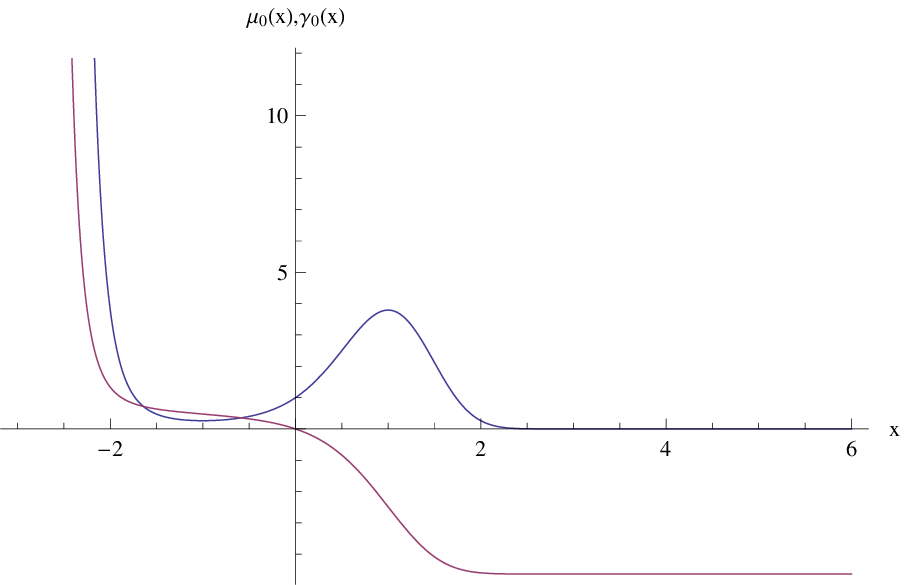}}}
\resizebox*{0.35\textheight}{!}{
{\includegraphics{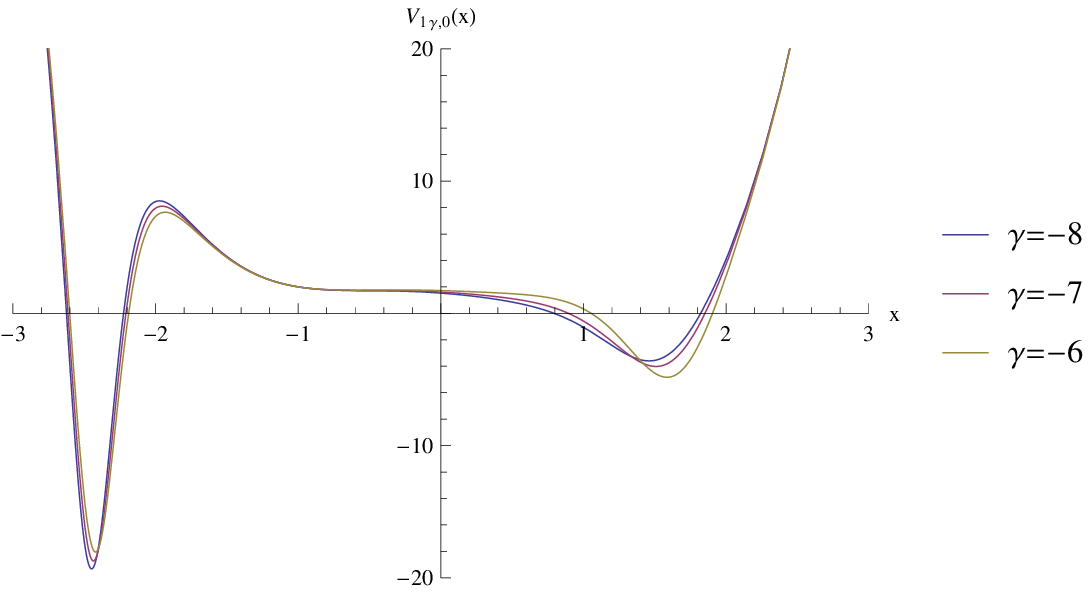}}}
\resizebox*{0.35\textheight}{!}{
{\includegraphics{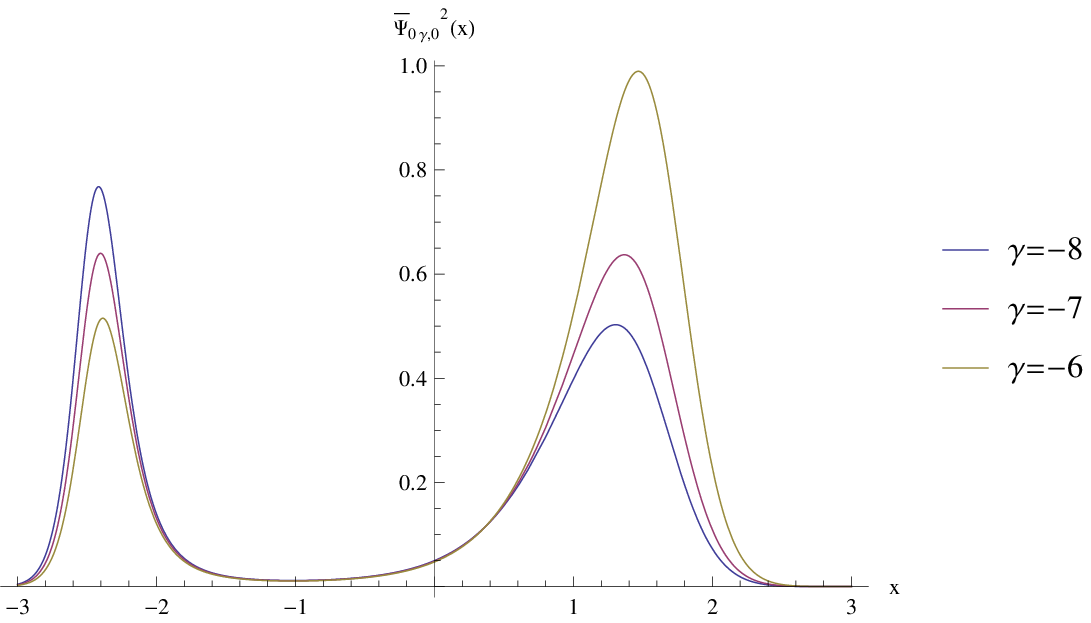}}}
\resizebox*{0.35\textheight}{!}{
{\includegraphics{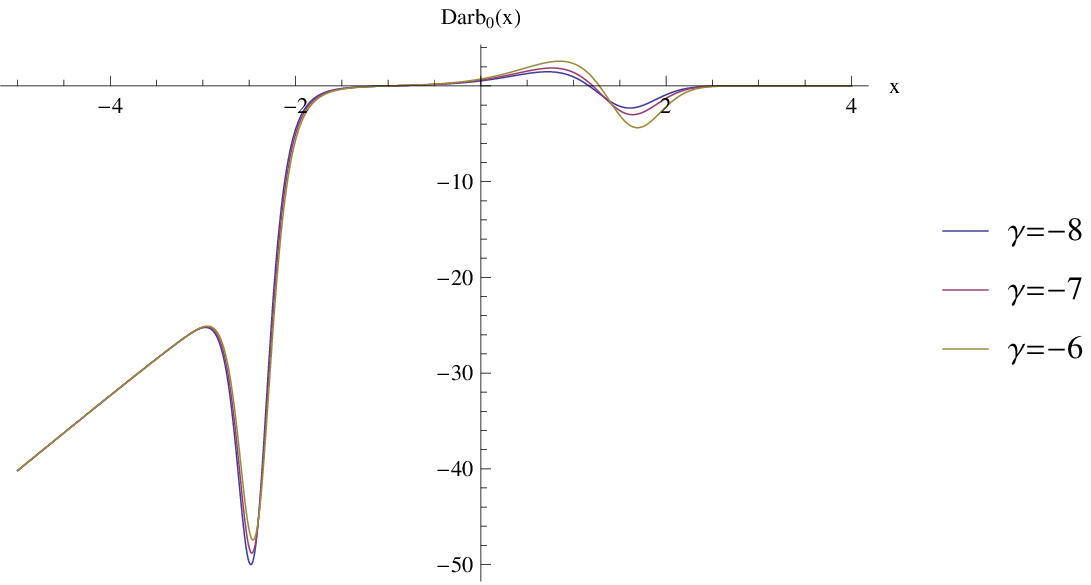}}}
\resizebox*{0.3\textheight}{!}{
{\includegraphics{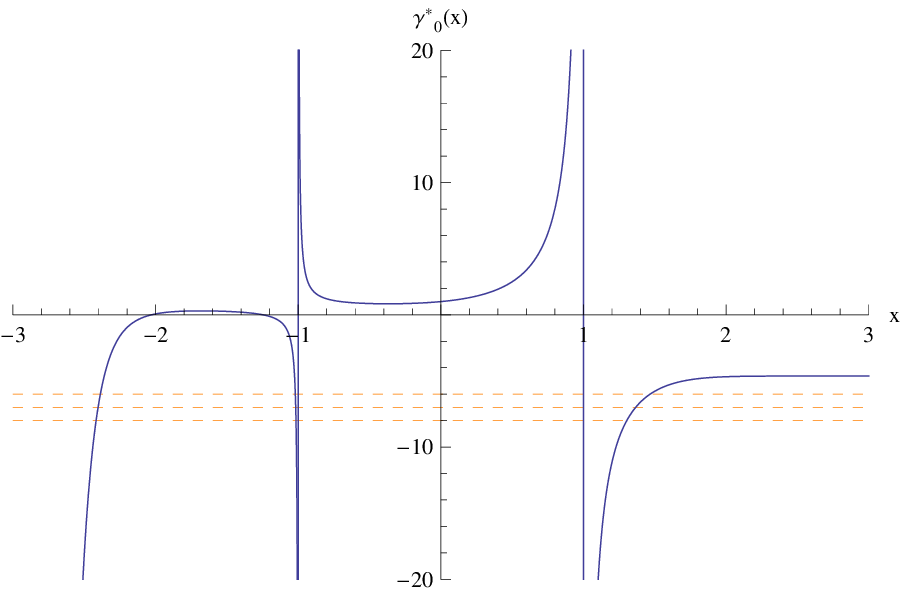}}}
\caption{\textsl{
Case $c=0$: (first row, top) the partner asymmetric quartic potentials, $V_1$ (blue) and $V_2$ (red) and the  integrating factor $e^{-\frac{2}{3}x(x^2-3)}$ (blue) and $\gamma(x;-3)$ (red); (second row) the one-parameter family of DWs, $V_{1\gamma}$ and the   squared normalized ZM eigenfunctions, $\bar{\Psi}_{0\gamma,0}^{2}$, for $\gamma=-8,-7$, and $-6$; (third row) the parametric Darboux deformation and the plots of $\gamma^*(x)$. There are two intersections with constant-$\gamma$ horizontal lines which provide the locations of the two peaks in the wells as given in Table 1. 
}}
\label{Set1}
\end{center}
\end{figure}

Plots of all these functions, of which the deformed ZMs with the additional normalization factor $N_\gamma=\sqrt{\gamma(\gamma+1)}$ \cite{rmc}, are presented in Fig.~\ref{Set1}. For the same unshifted case, the value of the threshold $\gamma_s$ beyond which on the negative vertical axis there are no intersections with the integral $\gamma(x)$ is given by
\begin{equation}\label{gs1}
\gamma_s=-\left({}_1F_2\left(1;\frac 4 3,\frac 5 3; \frac 4 9\right)+\frac \pi 3 2^{\frac 2 3} \mathrm{Bi}(2^{\frac2 3})\right) \approx -4.63107~,
\end{equation}
where $\mathrm{Bi}(x)$ is the Airy function of the second kind, whereas the value of $\gamma_{cr}$ is given by $\gamma_{cr}=-7$.
In the second plot of the first row of Fig.~\ref{Set1}, we display $\mu(x)$ and $\gamma(x)=-\int_{0}^{x}\mu_F(x')dx'$.
We notice graphically that the horizontal line $\gamma=const$ intersects $\gamma(x)$ for any given $\gamma >\gamma_s$, which generates singular potentials and wavefunctions, see the bottom row of Fig.~\ref{Set1} for such a case. Thus, only the range $\gamma<\gamma_s=-4.63107$ provides regular parametric potentials and normalized zero modes denoted by $\bar{\Psi}_{0\gamma}$ and differing from $\Psi_{0\gamma}$ by the normalization factor $\sqrt{\frac{\gamma(\gamma+1)}{|\Gamma|}}$, where we have used $|\Gamma|=|-\int_{l}^{\infty}\mu_F(x')dx'|=17.56$ obtained with the lower limit $l=-2.425$.

\bigskip

\begin{figure}[x!] 
\begin{center}
\resizebox*{0.35\textheight}{!}{
{\includegraphics{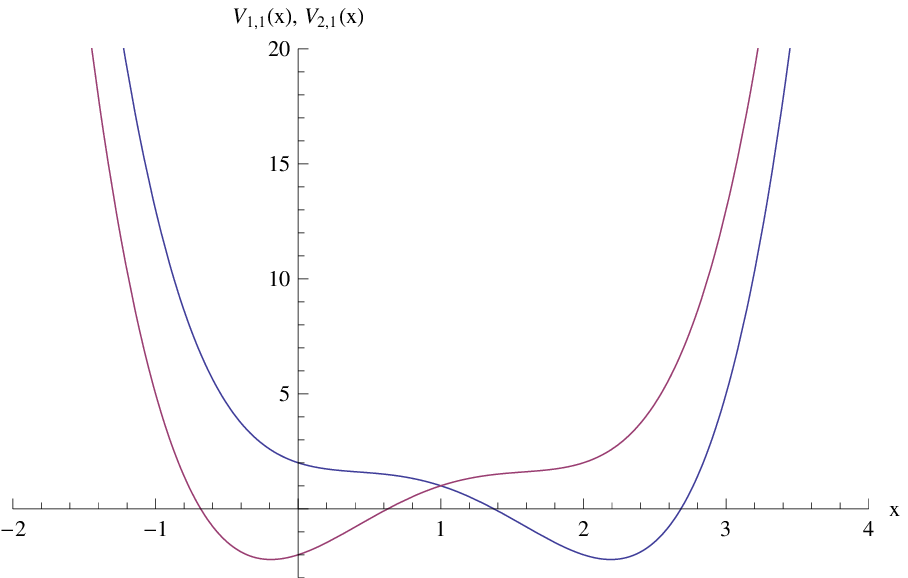}}}
\resizebox*{0.35\textheight}{!}{
{\includegraphics{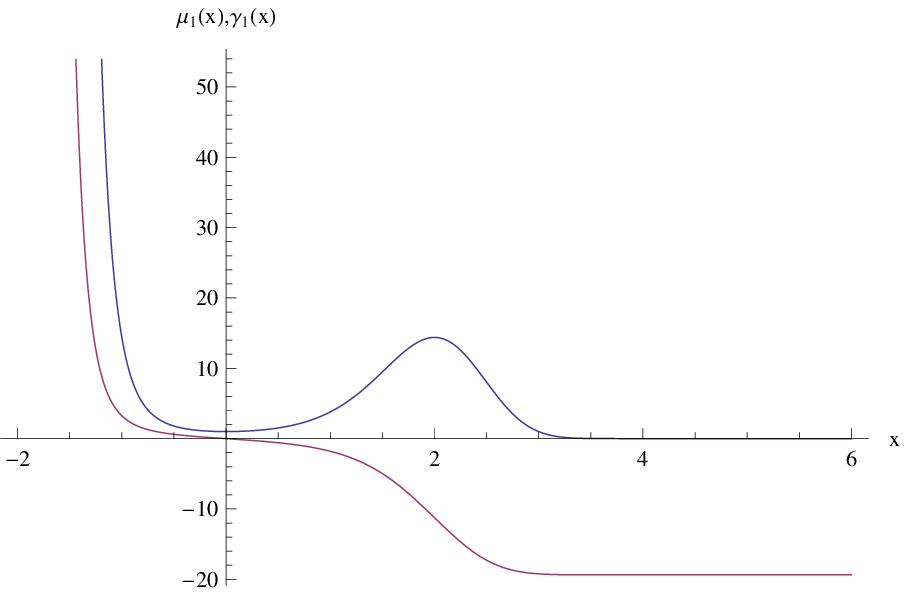}}}
\resizebox*{0.35\textheight}{!}{
{\includegraphics{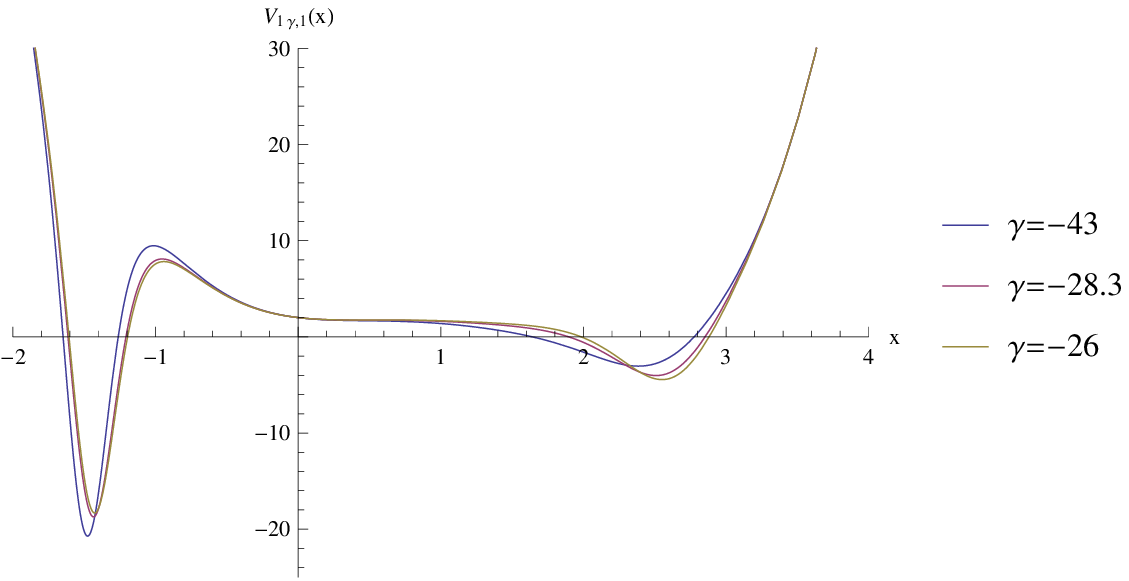}}}
\resizebox*{0.35\textheight}{!}{
{\includegraphics{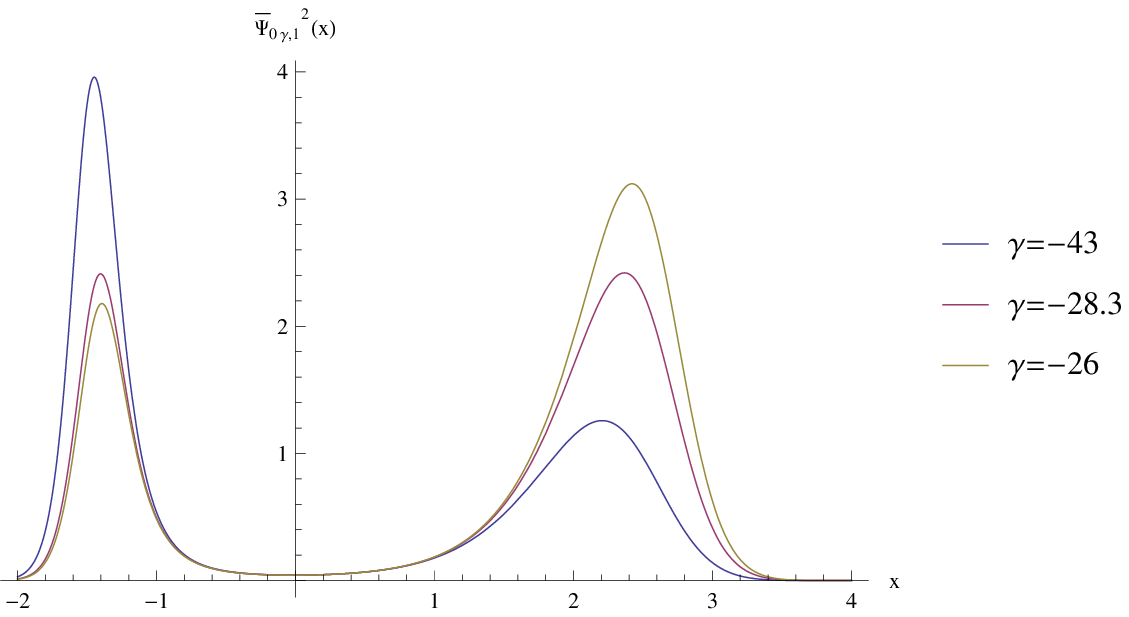}}}
\caption{\textsl{
Case $c=1$: the partner asymmetric quartic potentials, $V_1$ (blue) and $V_2$ (red);  integrating factor $\mu(x)$ (blue) and $\gamma(x)$ (red);   one-parameter family of potentials, $V_{1\gamma}$; squared ZMs $\bar{\Psi}_{0\gamma,1}^{2}$, for $\gamma=-43,-28.3$, and $-26$.}}
\label{Set12}
\end{center}
\end{figure}

Although we did many plots corresponding to the shifted cases, including $\gamma_c(x)$ and the normalized $\Psi_{0\gamma}(x)$, we present for illustrative purposes only the case $c=1$ in Fig.~\ref{Set12}.

\medskip

In Table 1, we provide the threshold values $\gamma_s$ and the critical value $\gamma_{cr}$ defining the borders of the ALRs, and also the locations of the peaks and minima of the ZM density distributions for all the cases presented in this work.
All the reported values have numerical errors $\leq$ 1\%.

\begin{center}
\begin{tabular}
[c]{|c|c|c|c|c|c|}\hline
\multicolumn{6}{|c|}{Table 1: Parameters of the quartic DWs and zero modes \ }\\\hline
{\small Shift} & {\small Threshold $\gamma$} & {\small Critical $\gamma$} & {\small $\gamma_{cr}$-ZM} & {\small $\gamma_{cr}$-ZM} & {\small $\gamma_{cr}$-ZM}\\
$c$ & $\gamma_s$ & $\gamma_{cr}$ & {\small left max.\ {\it x}} & {\small right max.\ {\it x}} & {\small local min.\ {\it x}}\\\hline
-2 & -0.1416 & -9.1 & -4.4 & -0.63 & -3.11\\ \hline
-1 & -0.5648 & -1.2 & -3.4 & 0.35 & -2.11\\ \hline
0 & $-4.6310$ & $-7.0$ & $-2.4$ & 1.36 & -1.02\\\hline
1 & $-19.3694$ & $-28.3$ & -1.4 & 2.35 & -0.10\\\hline
2 & $-1.5719$ & $-2.2$ & -0.4 & 3.35 & 0.91\\\hline
\end{tabular}
\end{center}

\bigskip

 Very useful information can be obtained from a careful examination of the plots included in this work. We have basically an initial potential $V_1$ with one left shoulder and a right shallow well that we deform to strongly asymmetric DWs through parametric Darboux deformations for different values of the parameter when the initial potential is shifted to various positions along the horizontal axis.  We first notice that the effect of the parametric Darboux deformations on the well of $V_1$ is a reduction of width in the region opposite to its right wall. On the other hand, the left deep well is produced by the left sharp negative peak of the Darboux deformation and its interplay with the opposite asymptotic behavior of the left wall of $V_1$.

 \medskip

As for the zero modes, the right bump in the plot of the integrating factor $\mu(x)$ is responsible for the part localized in the right shallow well since $\sqrt{\mu}$ is the undeformed ZM. The left peak is produced through a compensation effect between the upgoing plots of the integrating factor and its integral $\gamma(x)$ just before the latter one gains over the first and puts the ZM to naught because of its location in the denominator.
Another interesting effect occurs for $c\geq 0$ when $\gamma(x)$ becomes negative in the range of the $\mu$ bump. Then the right peak in the regular localization region is always smaller than the right peaks corresponding to the parameters within the ALR, whereas for $c<0$ the effect is opposite.

\medskip

When one goes further on with fixed negative shifts, one finds that an inversion in the deepness of the Darboux deformation occurs after $c\leq -2.6$.
Then, the parametric DW potentials have the right well deeper than the left one. There is even the possibility within numerical errors to get an almost symmetric DW parametric potential at finite values of the shift parameter and big but still finite absolute values of the parameter $\gamma$. On the other hand, for big positive values of the shift parameters, the left sharp well never crosses to the positive semiaxis and it is compressed towards the origin, while the right well becomes shallower and shallower in its rightward propagation.


 \medskip

The intersections of the horizontal lines $\gamma^*=const$ with $\gamma^*(x)$ are important not only because they can be used to determine $\gamma_{cr}$ but because they also provide the location of local extrema for $\bar{\Psi}_{0\gamma}$. For example, in the unshifted case when $\gamma^*=-7$ then $x_1=-2.404$ and $x_2=1.365$ are local maxima for $\bar{\Psi}_{0 \gamma}^2$, while  $x_3=-1.02$ is a local minimum. 
Taking into account this ZM minimum, then the localization probability in the left well is $\int_{-3}^{-1.02}\bar{\Psi}_{0\gamma}^2dx=0.31954$, while in the right well it is  $\int_{-1.02}^{3}\bar{\Psi}_{0\gamma}^2dx=0.68989$.
In addition, we notice another interesting feature: the roots of the potential $V_2$ at $x_1=-1.6837$ and $x_2=-0.3715$ are actually local maximum and minimum, respectively, for $\gamma^*$. 

\medskip

Regarding the singular parametric potentials, we address the reader's attention to the bottom row of Fig.~\ref{Set1}. One can see that for $\gamma>\gamma_s$ but still negative a split of the parametric potential into two branches occurs in the region of the right shallow well as if that well blows up while the left well does not disappear but becomes smaller and smaller. For positive $\gamma$'s the effect is opposite, i.e., the left well explodes and two separate branches of the parametric potential are created there, whereas the right well flattens out.

\bigskip

\section{A Razavy case}

\begin{figure}[x!] 
\begin{center}
\resizebox*{0.35\textheight}{!}{
{\includegraphics{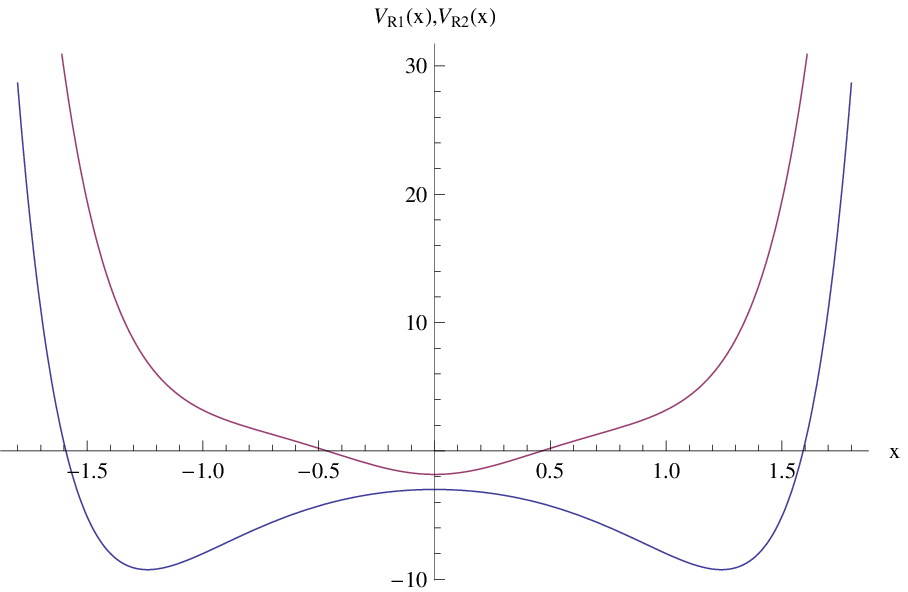}}}
\resizebox*{0.35\textheight}{!}{
{\includegraphics{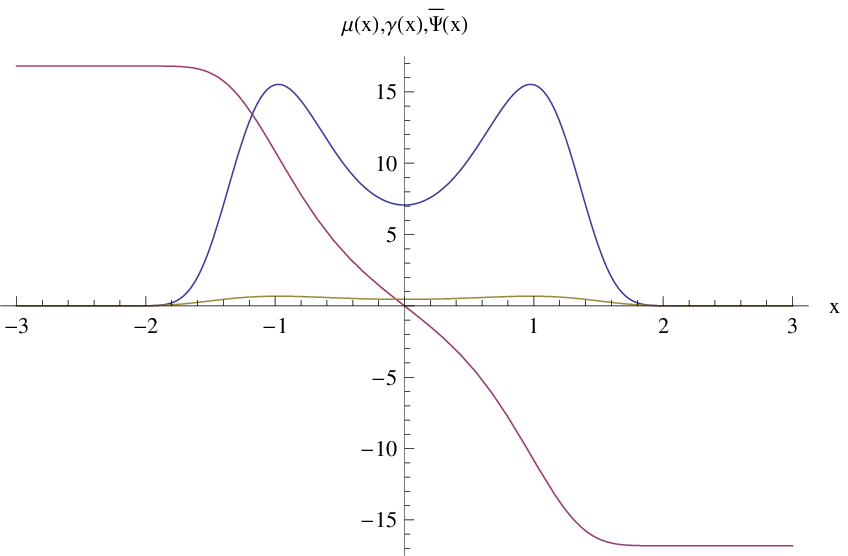}}}
\resizebox*{0.35\textheight}{!}{
{\includegraphics{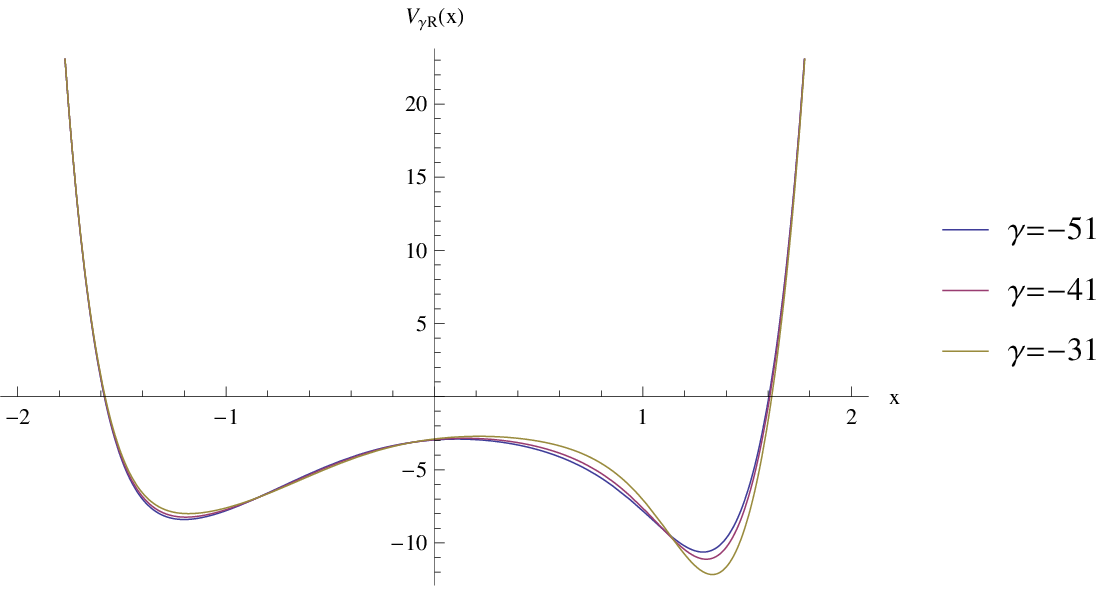}}}
\resizebox*{0.35\textheight}{!}{
{\includegraphics{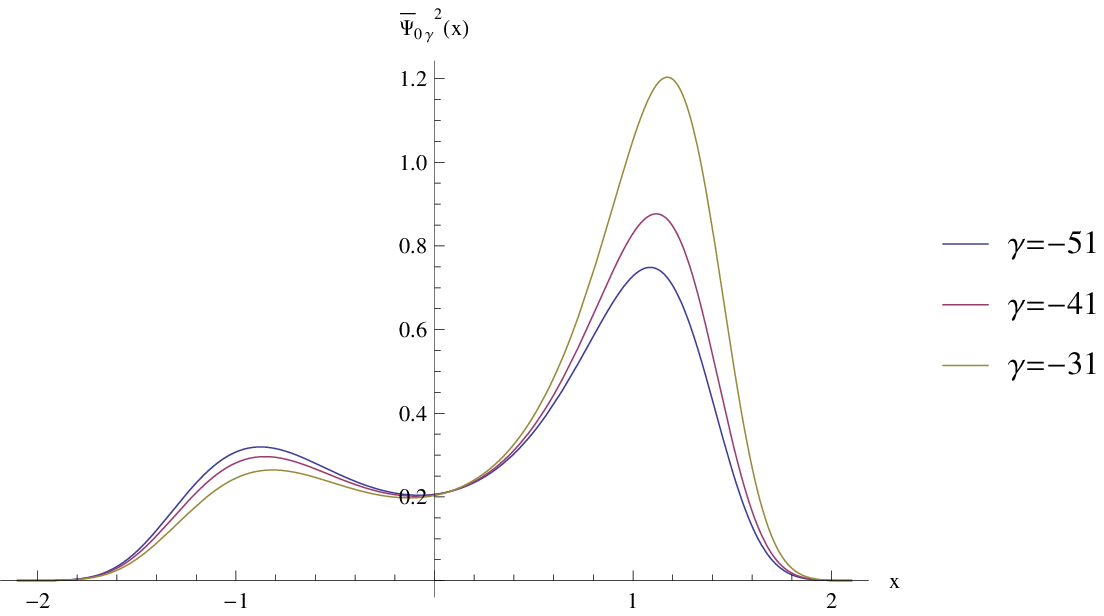}}}
\resizebox*{0.35\textheight}{!}{
{\includegraphics{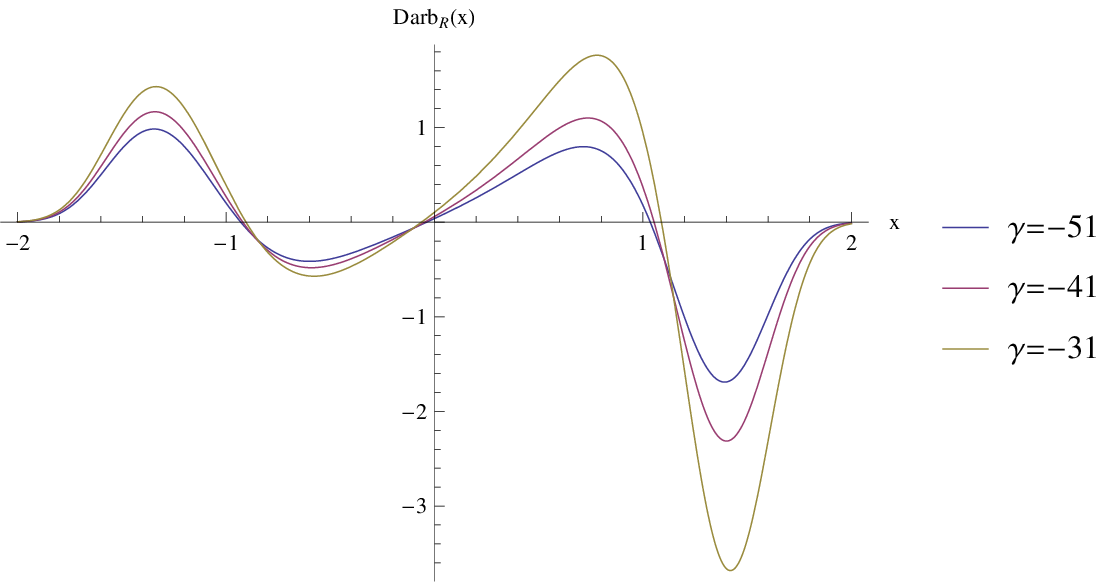}}}
\resizebox*{0.35\textheight}{!}{
{\includegraphics{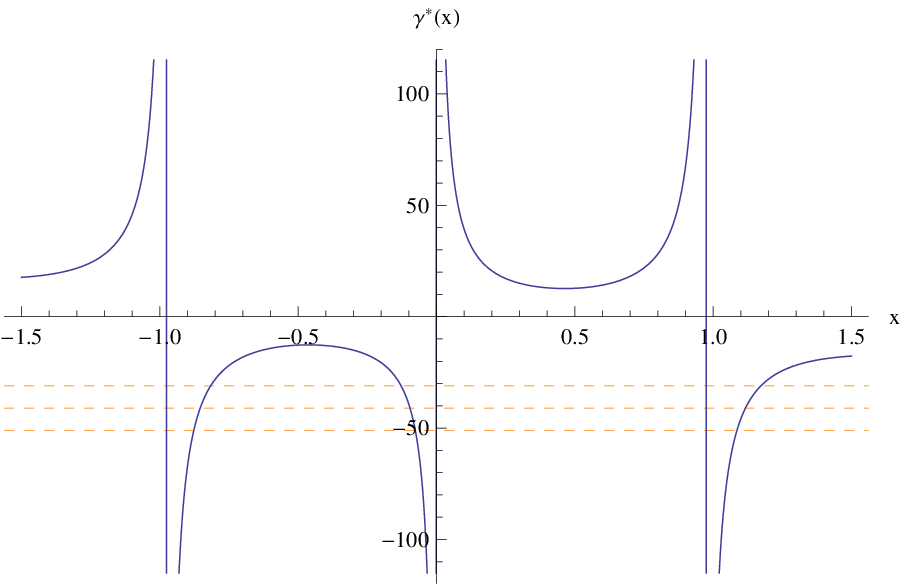}}}
\caption{\textsl{(Color online) Razavy's potential and supersymmetric partner for $n=2$, the integrating factor and its integral $\gamma(x)$ together with the undeformed normalized ZM, parametric isospectral potentials for three values of the parameter, and the corresponding normalized squared zero modes. Bottom row: the parametric Darboux deformation of $V_{R1}$ and the plot of $\gamma^*(x)$ for the same values of the parameter.
}}
\label{Raz1}
\end{center}
\end{figure}

For comparison, we check the localization properties for a similar double-well potential by applying the parametric scheme to one of Razavy's potentials with three parameters, $\beta$, $\xi$, and $n$, \cite{Raz80}
\begin{equation}\label{z1}
V_{R_n}(x)=\frac{\hbar^2\beta^2}{2m}\bigg[\frac{1}{8}\xi^2\cosh 4\beta x-(n+1)\xi\cosh 2\beta x-\frac{1}{8}\xi^2\bigg]~.
\end{equation}
Since the example is illustrative, we can fix $\xi=1$ and $\beta=1$, and we also take $\hbar=1$ and $2m=1$. Then, for $n=2$, we have
\begin{equation}\label{z2}
V_{R_2}(x)=\frac{1}{8}\cosh 4x-3\cosh 2x-\frac{1}{8}~,
\end{equation}
which is known to be an exactly solvable symmetric DW. Its plot, as well as that of its supersymmetric partner, are given in Fig.~\ref{Raz1}.
The Riccati solution which provides this potential is
\begin{equation}\label{riccraz}
F(x)=\frac{1}{2}\sinh 2x-2\frac{\epsilon\sinh 2x}{\epsilon \cosh 2x -2}~,
\end{equation}
where $\epsilon=-2(1+\sqrt 2)$.
According to Table I in Razavy's paper, the ground state solution of $V_{R_2}(x)$ is
\begin{equation}\label{z2p}
\psi_0=e^{-\frac{\cosh 2x}{4}}\bigg[1+(1+\sqrt{2})\cosh 2x\bigg],
\end{equation}
which solves the Schr\"odinger equation
\begin{equation}\label{z3p}
-\psi_0\rq{}\rq{}+(V_{R_2}(x)-\epsilon)\psi_0=0.
\end{equation}
The parametric potentials corresponding to Razavy's case $n=2$ are given by:
\begin{equation}\label{paramR}
V_{\gamma R}(x)=V_{R_2}(x)
-2(\ln\left |\gamma-\gamma(x)\right |)''~,
\end{equation}
where $\gamma(x)$ is the following integral:
\begin{equation}\label{gammax}
\gamma(x)=-\int_{0}^x e^{-\frac{\cosh 2x'}{2}}\bigg[1+(1+\sqrt{2})\cosh 2x'\bigg]^2dx'~.
\end{equation}
This integral is displayed in Fig.~\ref{Raz1} and shows a typical switching feature from a value of $\gamma_s \sim 16.8096$ to the opposite value $\gamma_s \sim -16.8096$.

The parametric ZMs are given by
\begin{equation}\label{parampsi}
\Psi_{0,\gamma}(x)=\frac{\psi_0(x)}{\gamma-\gamma(x)}~,
\end{equation}
where $\psi_0(x)$ is given by \eqref{z2p}.

\medskip

Plots of three parametric isospectral potentials for the Razavy case $n=2$ are given in Fig.~\ref{Raz1} and one can see that they are asymmetric double wells.
However, since the integrating factor $\mu_F$ is an even function with two symmetric equal peaks, we expect only zero modes showing a normal distribution between the two wells, i.e., lesser amplitude in the shallow well and more amplitude in the deeper well at moderate values of $\gamma$ and a saturation to a symmetric distribution in the two wells at higher values. This argument is validated by the rest of the plots in Fig.~\ref{Raz1}. In addition, the parametric Darboux transformations vanish beyond $x=\pm 2$ and have a left positive bump and a right deeper well for all negative values of the parameter in the regular region which explains the asymmetry of the parametric DWs in this Razavy case. Because $\mu(x)$ is an even function and $\gamma(x)$ has a common switching behavior no other effect is expected when the parametric potentials are shifted.

\medskip

As an example, we examine the intersections of $\gamma=-51$ with $\gamma^*$. They yield $x_1=-0.8783$ and $x_2=1.086$ as local maxima and $x_3=-0.076$ as the local minimum for $\bar{\Psi}_{0 \gamma}^2$, see the last plot in Fig.~\ref{Raz1}.
For normalization of the ZMs we calculate  $\Gamma=|\int_{-\infty}^{\infty}\mu dx|=33.6192=2|\gamma_s|$, because  the integrating factor $\mu(x)$ is even. Since the minimum of the ZMs is located  at $x=-0.076$, on the left we have $\int_{-3}^{-0.076}\bar{\Psi}_{0\gamma}^2dx=0.35311$, while on the right $\int_{-0.076}^{3}\bar{\Psi}_{0\gamma}^2dx=0.74674$.
When $V_{R_2}=0 $ then $x=\pm 0.463478$ are local maximum and minimum for $\gamma^*$, respectively, see the first and the last plots in Fig.~\ref{Raz1}.

\medskip

\begin{figure}[x!] 
\begin{center}
\resizebox*{0.35\textheight}{!}{
{\includegraphics{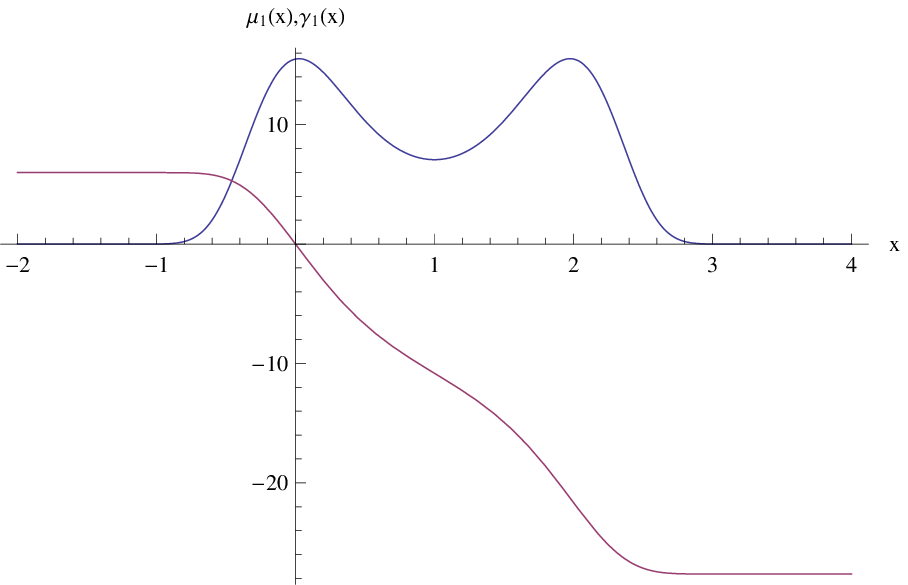}}}
\resizebox*{0.35\textheight}{!}{
{\includegraphics{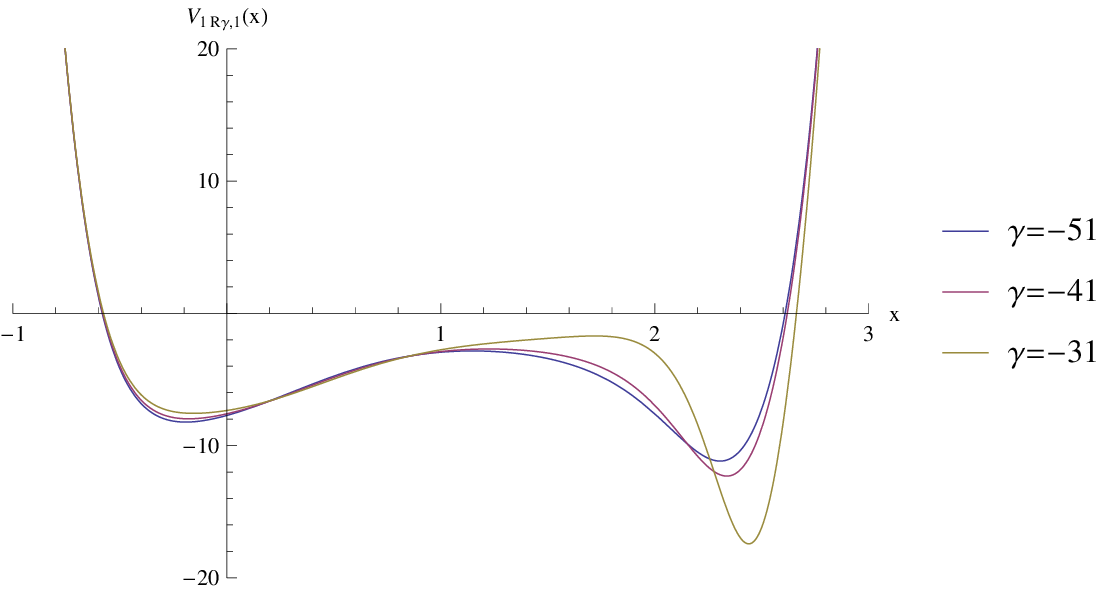}}}
\resizebox*{0.35\textheight}{!}{
{\includegraphics{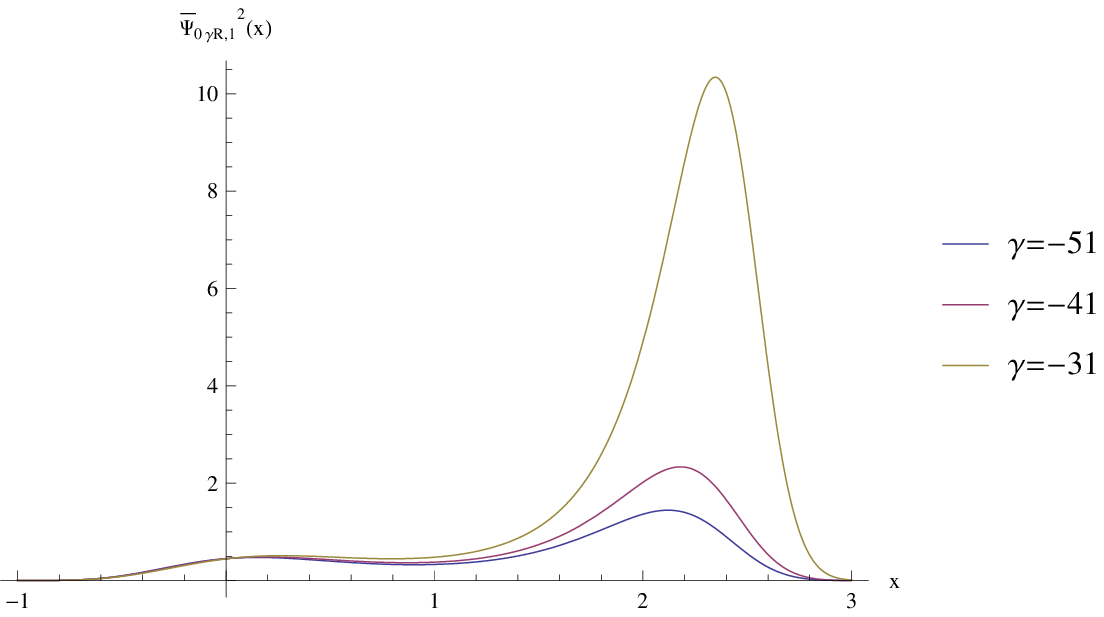}}}
\caption{\textsl{(Color online) The shifted $c=1$ case of $n=2$ Razavy parametric potentials and the corresponding normalized squared ZMs for three negative integer values of the $\gamma$ parameter. In the first plot, the shifted integrating factor and function $\gamma(x)$ are displayed.
For this case, $\gamma_s^+=5.986$ and $\gamma_s^-=-27.632$.}}
\label{Raz2}
\end{center}
\end{figure}

The one-parameter isospectral quantities obtained from the shifted Riccati solution of the Razavy type
\begin{equation}\label{riccrazc}
F(x-c)=\frac{1}{2}\sinh 2(x-c)-2\frac{\epsilon\sinh 2(x-c)}{\epsilon \cosh 2(x-c) -2}
\end{equation}
for $c=1$ are presented in Fig.~\ref{Raz2}. The switching (kink) shape of the integral function $\gamma(x)$ is preserved but the positive and negative plateau levels are not of the same absolute value which leads to changes in the threshold values of $\gamma_s$. If one goes to higher negative shifts, the positive $\gamma_s$ is bigger and bigger until it reaches the limiting value $33.6182$ and the negative $\gamma_s$ smaller and smaller in absolute value becoming naught in the asymptotic limit. For positive shifts, the negative $\gamma_s$ evolves until the limiting value of
$-33.6182$ and the positive one until naught asymptotically. We also considered other shift values without finding any ALR.

\section{Conclusions}

For different positions on the one-dimensional axis, we have studied in depth the phenomenon of anomalous heights of the square amplitude distribution of the zero modes of the quartic double well potentials in the class of parametric supersymmetric potentials generated from shifted quadratic Riccati solutions. When the parameter of these potentials is varied, there is a critical value at which the zero modes have higher peaks of the amplitude in the shallower well. It is known \cite{b90,m96} that the parameter of these supersymmetric potentials is related to the introduction of boundaries at certain finite points on the axis of the one-dimensional problem under study. Thus, varying this parameter is equivalent to moving boundaries and in the quartic case one will encounter the anomalous amplitude effect beyond a certain location of the distant boundary. In other words, one can produce this interesting effect by engineering distant boundaries. The Razavy case that we included for comparison shows no ALR feature. Thus the effect is not universal and depends on the shapes of the initial potential $V_1$ and integrating factor $\mu_F$.

\medskip

 The applications of the interesting effects found here should be carefully evaluated in those cases already discussed or suggested in the literature for other types of parametric isospectral potentials, i.e., to bound states in the continuum in quantum physics \cite{psp}, in photonic crystals \cite{pmr} and graded-index waveguides \cite{goyal}, as well as to generate soliton profiles \cite{kumar}. We also recall here the biological applications of harmonic oscillator isospectral potential to the simulation of the H-bond in DNA \cite{dfr} and traveling double-wells in microtubules \cite{rosuetal}. Another important application can be in the physics of Bose-Einstein condensates \cite{Dust}. However, as mentioned in the Introduction, the most direct application is in the form of effective potentials at the tip of a scanning tunneling microscope in the process of atom transfer during a voltage pulse \cite{bg98}. These effective potentials have been shown to be of the quartic type considered here and the scanning along one direction is equivalent to the shift of the coordinate considered by us. The parametric potentials can occur when the atomic transfer process is further constrained by boundaries of the scanned surface.

 \bigskip

{\bf Acknowledgment}: P.C. is supported in part by the Taiwan National Science Council under Project No. NSC 101-2923-M-
002-006-MY3 and 101-2628-M-002-006, by the Leung Center for Cosmology and Particle Astrophysics, National Taiwan University.

\renewcommand{\baselinestretch}{1.0}

\renewcommand{\baselinestretch}{1.0}

\renewcommand{\baselinestretch}{1.0}


\end{document}